%% file: Template.tex
\documentclass{article}
\usepackage{spconf,amsmath,graphicx,hyperref}
\usepackage{booktabs}
\usepackage{multirow}
\usepackage{subcaption}
\usepackage{colortbl} 
\usepackage{xcolor}   

\RequirePackage{xspace}
\makeatletter
\DeclareRobustCommand\onedot{\futurelet\@let@token\@onedot}
\def\@onedot{\ifx\@let@token.\else.\null\fi\xspace}

\def\eg{\emph{e.g}\onedot} 

\def\ie{\emph{i.e}\onedot}

\def\blue#1{\textcolor{blue}{#1}}
\definecolor{darkgreen}{RGB}{0,200,0}
\def\green#1{\textcolor{darkgreen}{#1}}
\makeatother

\usepackage{xcolor}
\newcounter{todos}
\AtEndDocument{\ifnum\value{todos}>0 \PackageWarning{TODOS}{There are \arabic{todos} todos left in this paper! Fix them before submitting the paper!} \fi}

\newcommand{\method}{AnimalCLAP\xspace}
\renewcommand{\green}[1]{\textcolor{black}{#1}}
\renewcommand{\blue}[1]{\textcolor{black}{#1}}
\title{AnimalCLAP: Taxonomy-Aware Language-Audio Pretraining \\ for Species Recognition and Trait Inference} 
\name{Risa Shinoda$^{1,2}$, Kaede Shiohara$^2$, Nakamasa Inoue$^3$, Hiroaki Santo$^1$, Fumio Okura$^1$ \thanks{This work was partly supported by JSPS KAKENHI JP25K24368 and JST FOREST JPMJFR206F.}}
\address{$^1$The University of Osaka\quad$^2$The University of Tokyo\quad$^3$Institute of Science Tokyo}
\begin{document}
%
\maketitle
\input{chapter/0_abst}
%
%

\input{chapter/1_intro}
\input{chapter/2_setup}
\input{chapter/3_method}

\input{chapter/4_experiment}

\input{chapter/5_conclusion}

\bibliographystyle{IEEEbib}
\bibliography{strings,refs}

\end{document}

%% file: chapter/0_abst.tex
\begin{abstract}
Animal vocalizations provide crucial insights for wildlife assessment, particularly in complex environments such as forests, aiding species identification and ecological monitoring.
Recent advances in deep learning have enabled automatic species \green{classification} from their vocalizations.
However, \green{classifying} species unseen during training remains challenging.
To address this limitation, we introduce \textbf{AnimalCLAP}, a taxonomy-aware language-audio framework comprising a new dataset and model that incorporate hierarchical biological information.
Specifically, our vocalization dataset consists of 4,225 hours of recordings covering 6,823 species, annotated with 22 ecological traits.
The AnimalCLAP model is trained on this dataset to align audio and textual representations using taxonomic structures, improving the recognition of unseen species.
We demonstrate that our proposed model effectively infers ecological and biological attributes of species directly from their vocalizations, achieving superior performance compared to CLAP. 
Our dataset, code, and models will be publicly available at \url{https://dahlian00.github.io/AnimalCLAP_Page/}.

\end{abstract}

%% file: chapter/1_intro.tex
\section{Introduction}
\vspace{-2mm}
\label{sec:intro}
\input{figure/teaser}

Automated recognition of animal vocalizations has emerged as an essential tool for biodiversity monitoring, particularly in visually complex habitats such as dense forests, where acoustic signals often provide the only reliable cues for species \green{identification}.
Traditionally, ecologists have relied on manual techniques, including field observations and spectrogram analyses, to document animal presence~\cite{Payne1971Humpback,monkeyresponse}.
Advances in acoustic sensing technologies, particularly automated recording units (ARUs), now facilitate large-scale continuous acoustic monitoring, highlighting the growing importance of sound analysis in ecological research~\cite{acoustic_monitoring_review, ecological_audio_review}.

Recent studies in signal processing have made substantial progress toward automating species \green{identification} from acoustic signals~\cite{Transferable,robinson2025naturelmaudio,inatsound,beans,Wood2022BirdNET}.
For example, BioLingual~\cite{Transferable} demonstrated the effectiveness of linking animal vocalizations to textual representations using contrastive language-audio pre-training (CLAP)~\cite{laionclap2023}, achieving impressive results in species \green{classification and detection tasks}.
NatureLM-Audio~\cite{robinson2025naturelmaudio} expanded the range of tasks by developing large-scale models that facilitate audio-based species retrieval.

Despite these successes, recognizing species unseen during training remains an open challenge.
Addressing this issue is crucial for building robust biodiversity monitoring systems because many species are inherently rare, making it difficult to collect sufficient training data.
This motivates us to highlight two key open questions.
The first question is how animal-specific textual knowledge can improve a joint audio-text feature space.
As animals are naturally organized into a hierarchical taxonomy, where names and categories reflect evolutionary and biological relationships among species, leveraging these hierarchical relationships could enhance the generalizability of audio-text representations.
While BioCLIP~\cite{stevens2024bioclip} has demonstrated that such hierarchical relationships can be effectively incorporated into image-text embeddings, this research direction has not yet been fully explored in the audio domain.
The second open question is
whether audio-text pre-training can connect animal vocalizations to ecological traits, such as habitat, diet, and activity patterns.
Although bioacoustics research has identified connections between animal vocalizations and environmental context~\cite{Morton1975Avian,Anthropogenic} or sociality~\cite{McComb2005Primate,social_complexity,Freeberg2006Chickadees}, these relationships remain unexplored within audio-text learning frameworks.

In this work, we introduce AnimalCLAP, a taxonomy-aware language-audio framework comprising a new dataset and model that incorporates hierarchical biological information.
Specifically, we collect animal vocalization recordings covering $6,823$ species, each annotated with taxonomic information and $22$ ecological traits, as shown in Figure~\ref{fig:teaser_animalclap}.
On this dataset, we train the AnimalCLAP model, which integrates the taxonomy structure into audio-text embeddings.
In the experiments, we evaluate how our training approach improves the \green{classification} performance for unseen species. Furthermore, we examine how detailed biological traits can be inferred from animal vocalizations. The results demonstrate the superiority of AnimalCLAP over the CLAP baseline. 

Our contributions are summarized as follows:
\vspace{-1mm}
\setlength{\leftmargini}{14pt}
\begin{enumerate}
\setlength{\parskip}{-3pt}
\item We construct the AnimalCLAP dataset, which consists of animal vocalizations from $6,823$ species annotated with $22$ trait labels. The dataset includes \blue{recordings from} rare species, \blue{making it a valuable} resource for audio-text learning and biodiversity monitoring.
\item We introduce the AnimalCLAP model, which leverages taxonomic structure during language-audio pre-training.
\item We demonstrate that our approach effectively generalizes to unseen species. As biological traits can be directly inferred from acoustic signals, our model maintains robust trait classification performance even for unseen species.
\end{enumerate}
\vspace{-3pt}

%% file: figure/teaser.tex
\begin{figure}[t]
\vspace{-2mm}
\centering
\includegraphics[width=1.0\linewidth]{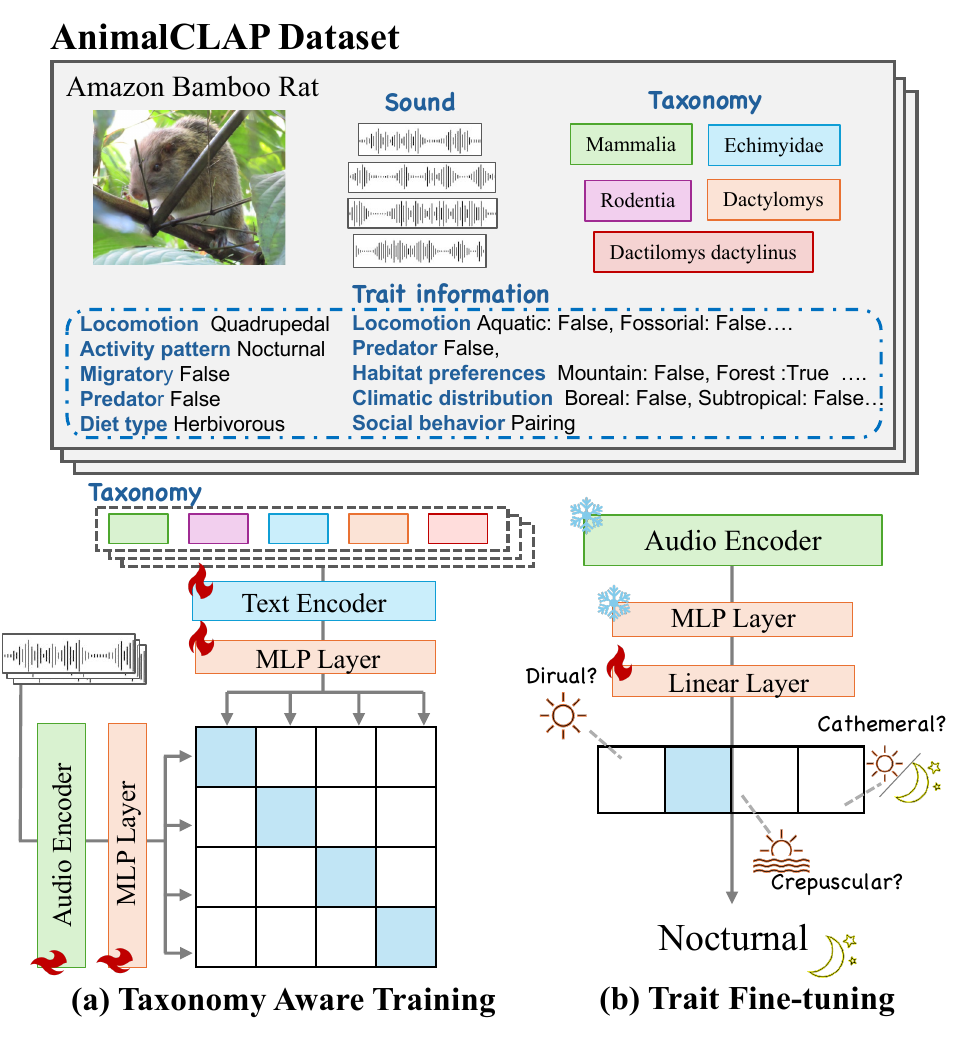}
\vspace{-23pt}
\caption{AnimalCLAP dataset and model. We newly introduce a taxonomy-aware training for language-audio pretraining.
}
\vspace{-13pt}
\label{fig:teaser_animalclap}
\end{figure}

%% file: chapter/2_setup.tex
\label{sec:setup}
\vspace{-3mm}
\section{\method Dataset}
\vspace{-2mm}

\label{sec:dataset}
The AnimalCLAP dataset consists of $4,225$ hours of animal vocalizations covering $6,823$ species.
Each audio recording is annotated with $22$ ecological trait labels.
For taxonomy-aware training and evaluation, we design three subsets for training, validation, and testing.
The test set comprises a carefully selected subset of 300 rare species, disjoint from those in the training and validation sets, for generalizability evaluation.

\vspace{-2mm}
\subsection{Dataset Construction}
\vspace{-2mm}
\input{tables/trait_list}

\input{tables/tax_ex}
\input{tables/bioclip}

\noindent \textbf{Data Collection.}
Audio recordings were collected from two platforms: iNaturalist~\cite{iNaturalist} and Xeno-canto~\cite{xenoCanto}. iNaturalist is a citizen-science platform where users submit biodiversity observations, including audio recordings of various species. We collected recordings uploaded to iNaturalist between 2014 and the first half of 2025. Xeno-canto is a community-driven repository primarily dedicated to bird vocalization recordings. We gathered recordings from Xeno-canto spanning the period from 2005 to the first half of 2025.

\noindent \textbf{Species Selection.}
We selected 6,823 species from the iNaturalist website~\cite{iNaturalist} that record ecological trait information.

\noindent \textbf{Trait Annotation.}
We defined 22 ecological traits for each species.
Table~\ref{tab:trait_list} summarizes the types and values of these traits, where categorical traits take one label from the provided values, and multi-label traits assign binary labels to each value.
Trait information was extracted from the iNaturalist website using GPT-5~\cite{OpenAI2025GPT5}. Extracted trait labels were subsequently verified manually, and missing information was completed accordingly.

\noindent \textbf{License.}
Only audio recordings published under Creative Commons licenses were included. Upon release, we will provide URLs for each audio recording and ensure compliance with their respective licenses.

\vspace{-3mm}
\subsection{Dataset Splits}
\vspace{-2mm}

The dataset was divided into three subsets: training, validation, and test. A total of 300 species were selected for the test set based on two criteria:
{
\vspace{-6pt}
\setlength{\leftmargini}{14pt}
\begin{enumerate}
\setlength{\parskip}{0pt}
\item[i)] We prioritized less common species in our collected data. Specifically, only species with fewer than 15 recordings were eligible, ensuring minimal exposure during training.
\item[ii)] Species were sampled in a class- and order-balanced manner.
Also, we selected unseen species whose genera and families were represented in the training subset.
\end{enumerate}
}
This approach maintains taxonomic connections between seen and unseen species, facilitating the evaluation of cross-species generalization, as unseen species share higher taxonomic ranks with species in the training set. For the training and validation splits, we applied a 9:1 ratio, ensuring same-day recordings were not divided across subsets. We selected validation and test sets from iNaturalist with only research-grade observations.
The final dataset consists of approximately 700k recordings from 6,823 species (6 classes, 66 orders, 341 families, 2,152 genera), with  630k recordings in the training set, 67k in the validation set, and 1.2k in the test set. \looseness=-1

%% file: tables/trait_list.tex

\begin{table}[t]
\centering
\resizebox{0.5\textwidth}{!}{
\begin{tabular}{lll}
\hline
\textbf{Trait} & \textbf{Type} & \textbf{Values} \\
\hline
Diet type & Categorical & herbivorous / carnivorous / omnivorous / specialized \\
Activity pattern & Categorical & diurnal / nocturnal / crepuscular / cathemeral \\
Locomotion (mode) & Multi-label & arboreal, aquatic, terrestrial, fossorial, aerial \\
Locomotion (posture) & Categorical & quadrupedal / bipedal / other \\
Habitat & Multi-label & forest, grassland, desert, wetland, mountain, urban \\
Climatic distribution & Multi-label & tropical, subtropical, temperate, boreal, polar \\
Social behavior & Categorical & solitary / pairing / grouping / herding \\
Predator & Binary & True / False \\
Migratory & Binary & True / False \\
\hline
\end{tabular}
}
\vspace{-8pt}
\caption{Schema used for trait annotation. 
}
\vspace{-8pt}
\label{tab:trait_list}
\end{table}

%% file: tables/tax_ex.tex

\begin{table}[t]
\centering
\small
\resizebox{0.5\textwidth}{!}{
\begin{tabular}{llp{0.4\textwidth}}
\hline
\textbf{Template}  & \textbf{Example (\textquoteleft  Anianiau)} \\
\hline
Common Name (Com) & \textquoteleft Anianiau \\
Scientific Name (Sci) & Magumma parva \\
Taxonomic Sequence (Tax) & Aves Passeriformes, Fringillidae Magumma, Magumma Parva \\
Sci + Com & Magumma Parva with a common name \textquoteleft Anianiau \\
\multirow{2}{*}{Tax + Com} & Aves Passeriformes, Fringillidae Magumma, Magumma Parva, \\
& with a common name \textquoteleft Anianiau \\
\hline
\end{tabular}
}
\vspace{-8pt}
\caption{Examples of textual descriptions following the five templates used in training.}
\vspace{-8pt}
\label{tab:tax_ex}
\end{table}

%% file: tables/bioclip.tex

\begin{table*}[t]
\centering
\small
\setlength{\tabcolsep}{2pt} 
\begin{tabular}{l *{5}{>{\centering\arraybackslash}p{0.9cm}} | *{5}{>{\centering\arraybackslash}p{0.9cm}} | *{5}{>{\centering\arraybackslash}p{0.9cm}} }
\toprule
& \multicolumn{5}{c}{Top-1 accuracy~[\%]} & \multicolumn{5}{c}{Top-5 accuracy~[\%]} & \multicolumn{5}{c}{mAP (Top-5)} \\[-0.1em]
\cmidrule(lr){2-6} \cmidrule(lr){7-11} \cmidrule(lr){12-16}
\vspace{-0.1em}
Train$\downarrow$ Test$\rightarrow$ 
& \scalebox{0.92}{Com} & \scalebox{0.92}{Sci} & \scalebox{0.92}{Tax} & \scalebox{0.86}{\hspace{-3pt}Sci\scalebox{0.9}{+}Com} & \scalebox{0.8}[0.86]{\hspace{-2pt}Tax\scalebox{0.9}{+}Com}
& \scalebox{0.92}{Com} & \scalebox{0.92}{Sci} & \scalebox{0.92}{Tax} & \scalebox{0.86}{\hspace{-3pt}Sci\scalebox{0.9}{+}Com} & \scalebox{0.8}[0.86]{\hspace{-2pt}Tax\scalebox{0.9}{+}Com}
& \scalebox{0.92}{Com} & \scalebox{0.92}{Sci} & \scalebox{0.92}{Tax} & \scalebox{0.86}{\hspace{-3pt}Sci\scalebox{0.9}{+}Com} & \scalebox{0.8}[0.86]{\hspace{-2pt}Tax\scalebox{0.9}{+}Com}\\
\midrule
Com     & \underline{19.5} & 0.81 & 2.33 & 14.6& 9.75 & \underline{42.1} & 1.97 & 6.71 & 33.5 & 25.5 & \underline{28.1} & 1.23 & 4.13 & 21.4 & 15.7 \\
Sci     & 0.36 & \underline{23.6} & 3.22 & 22.1 & 2.42 & 1.52 & \underline{50.7} & 11.4 & 49.8 & 8.77 & 0.70 & \underline{33.2} & 6.15& 32.2 & 4.59\\
Tax    & 1.16 & 3.49 & \underline{25.1} & 1.61 & 20.4 & 5.81 & 10.1 & \underline{51.5} & 4.11 & 49.6 & 2.90 & 5.76 & \underline{35.6} & 2.50 & 31.4 \\
Sci+Com & 14.3 & 10.6 & 4.20 & \underline{24.9} & 16.2 & 39.5 & 22.6 & 9.57 & \underline{50.4} & 37.7 & 23.6 & 14.7 & 6.24& \underline{34.5} & 24.1\\
Tax+Com & 7.60 & 1.43 & 19.7 & 7.51 & \underline{25.6} & 18.3 & 6.26 & 42.5 & 17.9 & \underline{53.1} & 11.4 & 2.88 & 28.1 & 11.0 & \underline{35.6}  \\
\rowcolor{blue!10} AnimalCLAP & \textbf{21.4} & \textbf{26.1} & \textbf{26.6} & \textbf{26.9} & \textbf{27.6} & \textbf{43.0} & \textbf{52.5} & \textbf{52.1} & \textbf{53.2} & \textbf{53.5} & \textbf{29.4} & \textbf{35.7} & \textbf{36.2} & \textbf{36.8} & \textbf{37.6}  \\[-0.2em]
\midrule
CLAP & 1.16 & 0.36 & 0.63 & 1.70 & 1.61 & 3.76 & 3.13 & 2.86 & 5.81 & 5.19 & 2.07 & 1.29 & 1.40 & 3.04 & 2.73 \\
\bottomrule
\end{tabular}
\vspace{-5pt}
\caption{Zero-shot accuracies on species not seen during training. The mean average precision (mAP) was computed based on the top-5 correctness. \textbf{Bold} and \underline{underline} indicate the best and second-best scores.}
\vspace{-8pt}
\label{tab:taxo_zero_acc}
\end{table*}

%% file: chapter/3_method.tex
\section{\method Model}
\vspace{-3pt}

The AnimalCLAP model learns to align audio and text representations in a joint embedding space, using taxonomic structure to enhance the generalization ability for unseen species. 

\vspace{-3pt}
\subsection{Taxonomy-Aware Pre-training}
To incorporate taxonomy information into audio-text embeddings, we train the CLAP model with prompts augmented by 1) Common name (Com), 2) Scientific name (Sci), and 3) Taxonomic sequence (Tax).
Specifically, given a training dataset $\mathcal{D} = \{ (x_{i}, y_{i} )\}_{i=1}^{N}$ consisting of audio clips $x_{i}$ paired with species labels $y_{i}$, we compute the similarity between audio and text embeddings as follows:
\begin{align}
s_{ij} = \exp(\gamma) \, \frac{f_{a}(x_{i}) \cdot f_{t} ( \phi(y_{j}) )}{\|f_{a}(x_{i})\| \, \|f_{t} (\phi(y_{j}))\|},
\end{align}
where $f_{a}$ is an audio encoder, $f_{t}$ is a text encoder, $\phi$ is an augmentation function and $\gamma$ is a hyperparameter.
The augmentation function $\phi$ randomly selects one of the five prompts (Com, Sci, Tax, Sci+Com, and Tax+Com) defined in Table~\ref{tab:tax_ex}.
For instance, given the species \textit{Anianiau},
the augmented prompts include the scientific name (\textit{Magumma parva}) and its taxonomic order (\textit{Aves Passeriformes}).
The model is trained using the CLIP contrastive loss~\cite{CLIP} to maximize the similarity for correct pairs and minimize it for incorrect pairs.
This strategy encourages a robust alignment between audio and textual representations in a structured manner.

\noindent
\textbf{Implementation Details.}
Audio recordings were resampled to 48 kHz and randomly cropped into 10-second clips. 
We adopted the encoders used in CLAP~\cite{laionclap2023}: HTS-AT~\cite{htsatke2022} as the audio encoder, and RoBERTa-based Transformer~\cite{liu2019robertarobustlyoptimizedbert} as the text encoder.
Two-layer MLP heads were added on top of these encoders as shown in Figure~\ref{fig:teaser_animalclap}(a).
The model was trained for 20 epochs using AdamW~\cite{loshchilov2018decoupled} 
with a learning rate of $10^{-4}$, where one epoch corresponds to a pass over a balanced dataset constructed by sampling 30 clips per species.

\vspace{-3pt}
\subsection{Ecological Trait Fine-tuning}
For the 22 ecological traits we annotated, we fine-tuned the model to directly predict trait labels from sound representations. 
The architecture consists of an audio encoder followed by two MLP layers and a linear classifier, as shown in Figure~\ref{fig:teaser_animalclap}(b).
We initialized the encoder and MLP with pretrained weights and kept them frozen, training the linear classifier for~5 epochs with cross-entropy loss for multiclass traits and binary logistic loss for binary traits.

%% file: chapter/4_experiment.tex
\input{figure/tnse_clap}
\input{figure/tax}

\vspace{-3pt}
\section{Experiments}
\vspace{-3pt}
\label{sec:experiment}
\input{tables/trait}

We conducted experiments on the key scientific questions regarding our AnimalCLIP, \ie, the importance of taxonomic and hierarchical structures in our model and the prediction ability of ecological traits from audio information.

\vspace{-3pt}
\subsection{Does Taxonomy Improve Generalizability?}

To evaluate the effectiveness of language-audio training utilizing the taxonomic structure, we compare our AnimalCLAP model with the baseline CLAP as well as models trained exclusively on single prompt types. Table~\ref{tab:taxo_zero_acc} summarizes species classification accuracy on the test set. Across all metrics, the AnimalCLAP model consistently achieves the highest performance. Single-type models (\eg, Sci and Tax) excel in their respective query types but demonstrate weaker generalization across other types, whereas our proposed model sustains robust performance uniformly across all test settings.

The performance is also influenced by test prompts.
When queried with scientific names, AnimalCLAP scores $26.1$~\% top-1 accuracy, higher than $21.4$~\% achieved with common names. This suggests that scientific names, composed of genus and species, provide less ambiguous and more structured signals than culturally variable common names.\looseness=-1

Figure~\ref{fig:tsne} visualizes the embeddings obtained from the audio encoder on the validation dataset, using t-SNE with the six most frequent categories at each taxonomy level. In the top row, we observe that the AnimalCLAP model exhibits clearer embedding clusters aligned with the taxonomic hierarchy (class, order, family) compared to CLAP.

\vspace{-3pt}
\subsection{Is Biological Hierarchy Essential?}

To validate whether the hierarchical structure contributes to accuracy improvements, we tested the method where the order of elements within the taxonomic sequence (\ie, class, order, family, genus, species, scientific name) is randomized.
Table~\ref{tab:taxo_order_table} compares the top-1 accuracy between the ordered and randomized conditions.
Randomizing the taxonomic order significantly reduces top-1 accuracy across all test prompts, highlighting the importance of hierarchical structure.
The broad-to-narrow ordering (\ie, class $\rightarrow$ \dots $\rightarrow$ species) supports the learning of biological hierarchies, likely because the text encoder benefits from a coherent hierarchical sequence.\looseness=-1

Figure~\ref{fig:tax_order} presents an error analysis, showing the proportion of cases where the predictions for common names were incorrect at the species level but correct at higher taxonomic ranks (\ie, genus, family, order, and class).
The ordered condition yields substantially higher match rates from the class down to the genus level compared to the randomized condition, indicating that misclassifications are more taxonomically coherent when the training prompt follows a consistent hierarchical order.
These findings demonstrate that presenting taxonomic information in a broad-to-narrow sequence helps the model internalize biological hierarchies more effectively.\looseness=-1

\vspace{-3pt}
\subsection{Can Ecological Traits be Predicted?}
Table~\ref{tab:traits} shows the classification results of ecological traits for the test set.
Overall, our method consistently outperforms the CLAP baseline across all tasks, highlighting the feasibility of inferring diverse ecological traits directly from sound.

The improvement is particularly pronounced for behavioral traits such as Activity pattern, Locomotion, and Migration. 
These traits are closely linked to the temporal or locomotor characteristics of species, which are often reflected in their vocal behavior. 
For example, urban birds often shift song frequency to cope with noise, while vegetation and structures also influence acoustic adaptation~\cite{10.1093/beheco/arw105}.
As a result, these traits can be captured more directly by acoustic features, explaining the large performance gains.

In contrast, the performance gain is more modest for broad environmental traits such as forest in Habitat, and tropical or subtropical in Climate. 
One possible biological explanation is that these categories cover vast areas and encompass high ecological diversity. 
For instance, forests can host many different types of animals, such as birds and mammals, while tropical and subtropical zones include a wide variety of taxa. 
Because these categories contain multiple types of animals with heterogeneous vocal behaviors, their acoustic signatures are less consistent. 
Nevertheless, our results show that such traits can still be learned from acoustic data. 

Overall, these findings suggest that acoustic information provides a powerful tool for classifying species' behavioral and ecological strategies.

%% file: figure/tnse_clap.tex
\begin{figure}[t]
\centering
\includegraphics[width=1.0\linewidth]{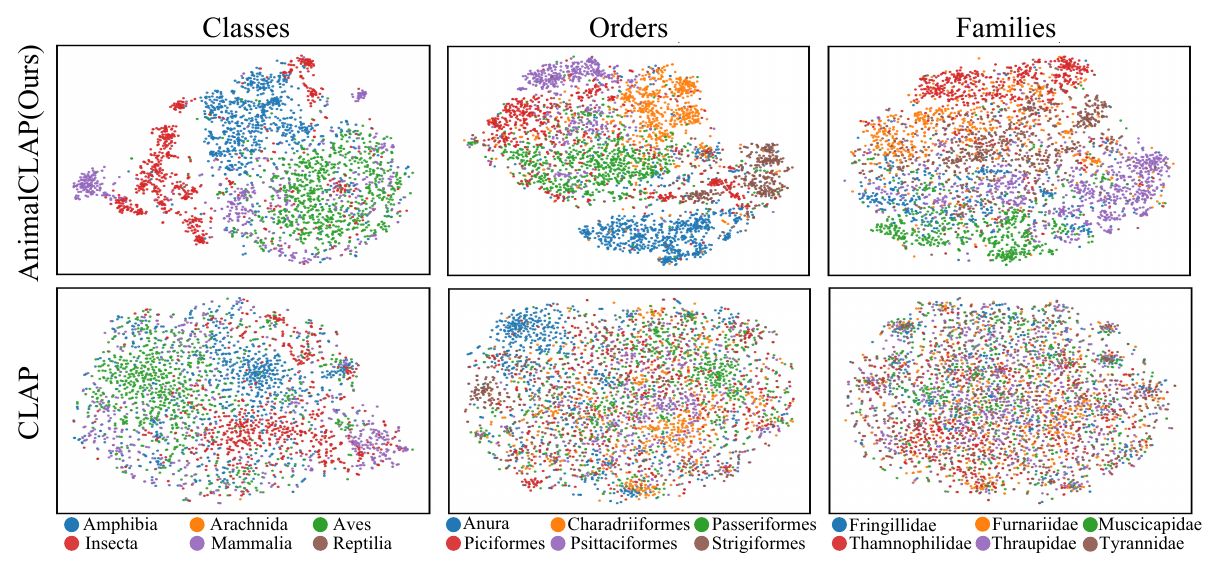}
\vspace{-20pt}
\caption{t-SNE visualization of AnimalCLAP and pretrained CLAP for validation data.}
\label{fig:tsne}
\end{figure}

%% file: figure/tax.tex
\begin{figure}[t]
  \centering
  \hspace{-5mm}
  \begin{minipage}[t]{0.48\linewidth}
    \vspace{0pt}%
    \centering
    \small
    \resizebox{\linewidth}{!}{%
      \begin{tabular}{lccc}
        \toprule
        \multirow{2}{*}{Tax order} & \multicolumn{3}{c}{Evaluation prompt} \\
        \cmidrule(lr){2-4}
         & Com & Sci & Tax \\
        \midrule
        Random & 19.8 & 21.3 & 22.5 \\
        Ordered  & 21.4 & 26.1 & 26.6 \\
        \bottomrule
      \end{tabular}
    }
    \captionof{table}{Top-1 accuracy with different taxonomic orders.}
    \vspace{-5pt}
    \label{tab:taxo_order_table}
  \end{minipage}
  \hspace{0.00mm}
  \begin{minipage}[t]{0.52\linewidth}
    \vspace{-3pt}
    \centering
    \includegraphics[width=\linewidth]{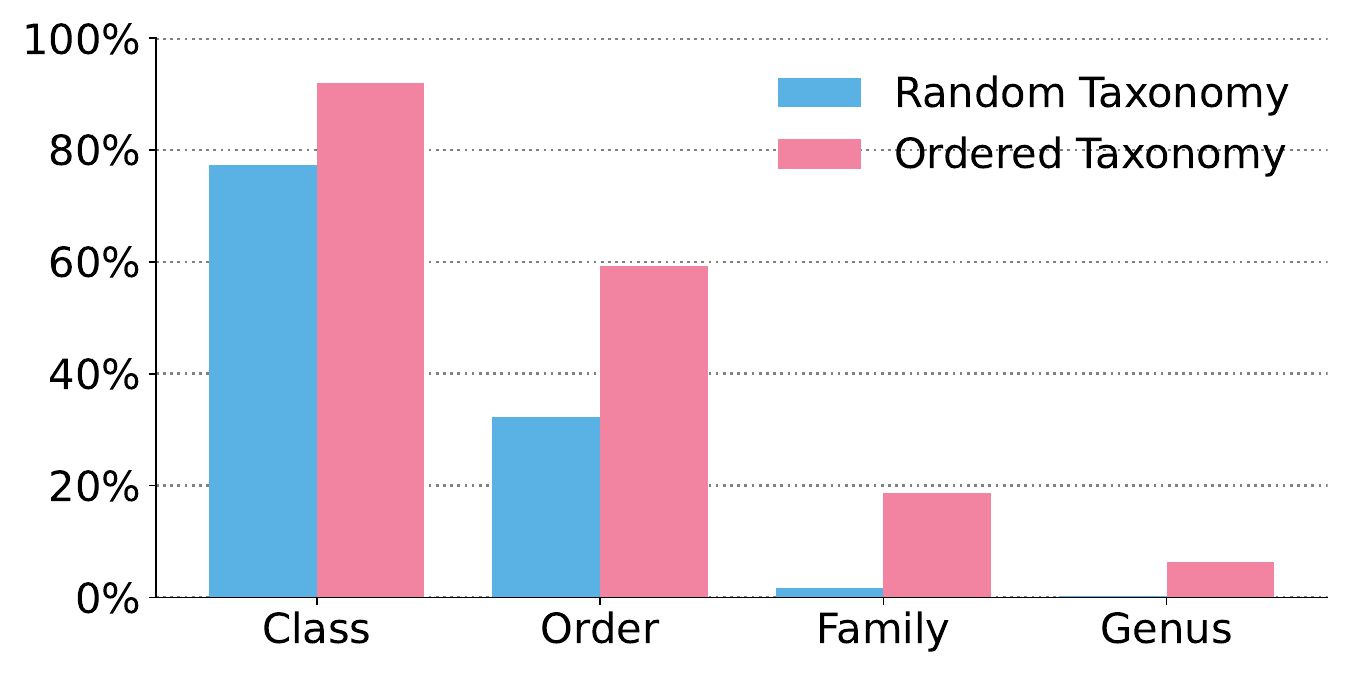}
    \vspace{-20pt}%
    \caption{\hspace{-1mm}Taxonomy accuracy with incorrect species prediction.}
    \vspace{-5pt}
    \label{fig:tax_order}
  \end{minipage}\hspace{-5mm}\hfill

\end{figure}

%% file: tables/trait.tex
\begin{table*}[t]
\centering
\small
\resizebox{\textwidth}{!}{
\begin{tabular}{lccccccccccc}
\toprule
\multirow{2}{*}{Method} &
\multirow{2}{*}{Diet type} &
\multirow{2}{*}{Activity} &
\multicolumn{5}{c}{Locomotion (dict)} &
\multirow{2}{*}{Loc. mode} &
\multirow{2}{*}{Social} &
\multirow{2}{*}{Pred.} &
\multirow{2}{*}{Migr.} \\[-0.2em]
\cmidrule(lr){4-8}
&  & 
& Arb. & Aquat. & Terr. & Foss. & Aerial
& \\[-0.1em]
\midrule
CLAP & 29.5 & 28.4 & 48.9 &83.3&38.2 & 72.2 &49.9 &4.17 & 26.2 & 29.8 &10.6\\
\rowcolor{blue!10} AnimalCLAP & \textbf{59.4} & \textbf{83.7} & \textbf{79.0} & \textbf{89.0} & \textbf{68.5} &  \textbf{92.6}& \textbf{84.8} & \textbf{87.5} &\textbf{52.6} &\textbf{87.2}&\textbf{84.0} \\[-0.2em]
\midrule
\midrule
\multirow{2}{*}{Method} &
\multicolumn{6}{c}{Habitat (dict)} &
\multicolumn{5}{c}{Climatic distribution (dict)} \\[-0.2em]
\cmidrule(lr){2-7}\cmidrule(lr){8-12}
& Forest & Grass. & Desert & Wetl. & Mount. & Urban
& Trop. & Subtrop. & Temp. & Boreal & Polar \\[-0.1em]
\midrule
CLAP & 76.1 & 35.3  &53.3 &29.2&46.1&48.2&73.0&60.6&53.5&60.7&73.8\\
\rowcolor{blue!10} AnimalCLAP & \textbf{81.7} & \textbf{69.9}& \textbf{88.4} &\textbf{63.2}&\textbf{59.8}&\textbf{72.3}&\textbf{83.0}&\textbf{64.4}&\textbf{76.6}&\textbf{90.5}&\textbf{98.5}\\[-0.2em]
\bottomrule
\end{tabular}
}
\vspace{-3pt}
\caption{Quantitative comparison of F1 scores between CLAP and Ours. \textbf{Bold} represents the best scores.}
\vspace{-8pt}
\label{tab:traits}
\end{table*}

%% file: chapter/5_conclusion.tex
\section{Conclusion}
\label{sec:conclusion}
We introduced \method model, which integrates the taxonomy structure into audio-text embeddings. 
Our experiments showed that hierarchical information improves the model’s ability to generalize to unseen species during training. 
In addition, our proposed \method dataset can serve as a new benchmark for trait prediction of unseen species.